\documentclass[aps,pre,preprint,superscriptaddress,showpacs]{revtex4}

\usepackage[utf8]{inputenc}
\usepackage{amsmath}
\usepackage{graphicx}
\usepackage{epstopdf}
\usepackage{color}

\begin{document}

\title{Phase of the transmitted wave in the dynamical theory and quasi-kinematical approximation}
\date{\today}

\author{O.Yu.~Gorobtsov}
	\affiliation{Deutsches Elektronen-Synchrotron DESY, Notkestra{\ss}e 85, D-22607 Hamburg, Germany}
	\affiliation{National Research Centre "Kurchatov Institute", Akademika Kurchatova pl., 1, 123182 Moscow, Russia}
\author{I.A.~Vartanyants}
\email[Corresponding author: ]{ivan.vartaniants@desy.de}
\affiliation{Deutsches Elektronen-Synchrotron DESY, Notkestra{\ss}e 85, D-22607 Hamburg, Germany}
\affiliation{National Research Nuclear University MEPhI (Moscow Engineering Physics Institute), Kashirskoe shosse 31, 115409 Moscow, Russia}

\begin{abstract}
Variation of the phase of the beam transmitted through a crystalline material as a function of the rocking angle is a well known dynamical effect in x-ray scattering.
Unfortunately, it is not so easy to measure directly these phase variations in a conventional scattering experiment.
It was recently suggested that the transmitted phase can be directly measured in ptychography experiments performed on nanocrystal samples.
Results of such experiment for different crystal thickness, reflections and incoming photon energies, in principle, can be fully described in the frame of dynamical theory.
However, dynamical theory does not provide a simple analytical expression for the further analysis.
We develop here quasi-kinematical theory approach that allows to describe correctly the phase of the transmitted beam for the crystal thickness less than extinction length that is beyond applicability of the conventional kinematical theory.

\end{abstract}


\maketitle

\section{Introduction}

It is well known that while x-rays are propagating through a slab of material a small phase is accumulated due to refraction \cite{Als-Nielsen}.
Being small, refraction coefficient for x-rays is on the order of $10^{-5}$ to $10^{-7}$, this phase plays an important role in different applications of x-rays.
For example, focusing of x-rays by compound refractive lenses \cite{Snigirev} and phase contrast imaging \cite{Wilkins} are based on that effect.
It is also well known from the dynamical theory \cite{Authier} that in the case of perfect crystal the phase of diffracted wave in Bragg geometry changes by $\pi$ in the narrow angular range close to Bragg angle.
This is a manifestation of x-ray standing wave generated in a crystal by two coherent transmitted and diffracted waves \cite{Batterman,VartKov,Standing_Wave}.
It is interesting to note that the phase of transmitted wave has also an additional dynamical correction close to Bragg angle that depends on the rocking angle.
Moreover, due to dynamical effects this phase correction depends on polarization of the incoming beam, being different for $\pi-$ and $\sigma-$ polarization of x-rays \cite{Dmitrienko,Belyakov}.
This effect is used nowadays to develop quarter and half-wavelength phase shifters for hard x-rays \cite{Hirano1991,Giles} and produce circular polarized radiation at 3-rd generation synchrotron sources.
Probably the most recent application of the transmitted beam phase in Bragg scattering geometry is generation of self-seeded pulses of x-ray free-electron lasers \cite{Geloni,Amann}.

\begin{figure}
	\includegraphics[width=\linewidth]{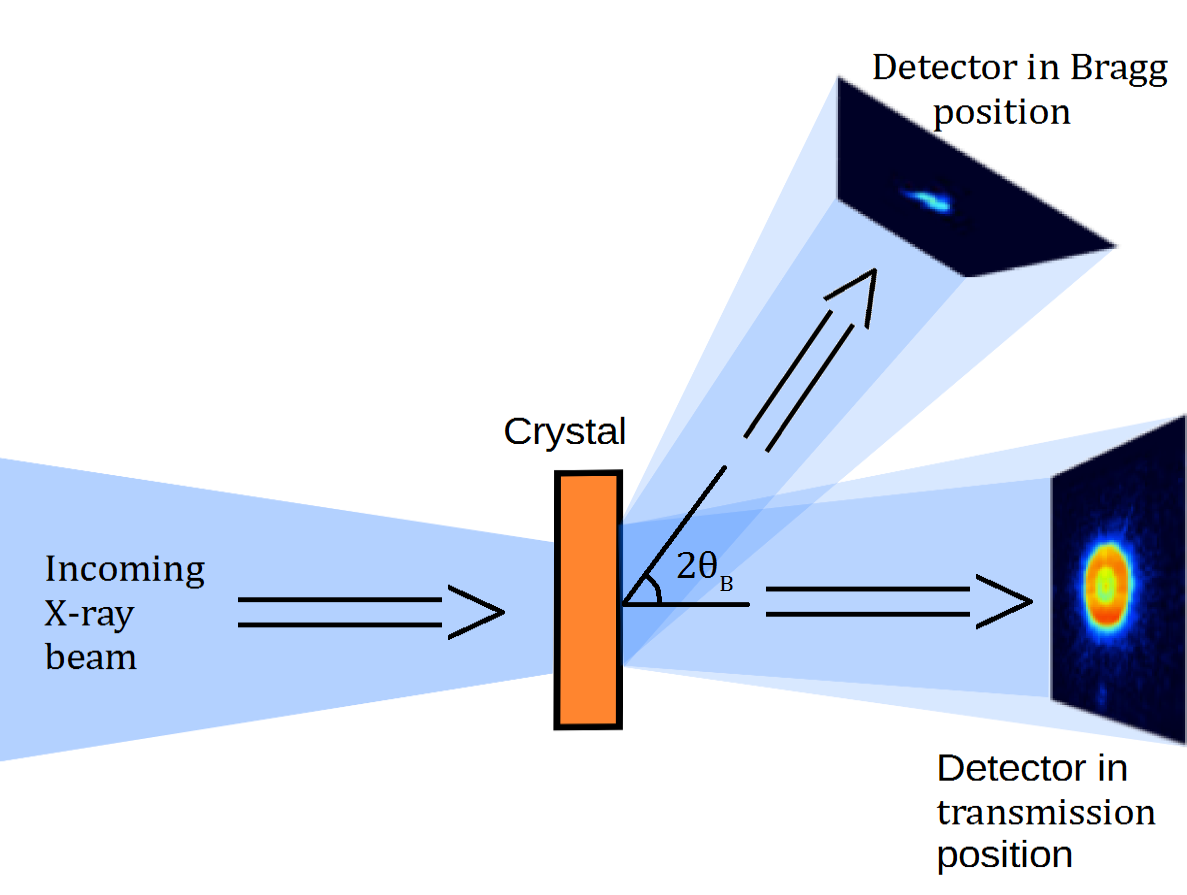}
	\caption{
    \label{fig::schematic}
		Schematic layout of the ptychography experiment described in Ref. \cite{Robinson}.
        Intensities of the transmitted and diffracted beams are measured simultaneously by two detectors, while the crystal is rotated near the Bragg angle.
        Ptychographic measurements were performed on a Au 111 crystalline nanoparticle 100 nm thick.
        Reconstruction of the complex amplitude of the transmitted beam allowed to observe phase variations as a function of the rocking angle.
	}
	\newpage
\end{figure}

It is not so easy to measure the phase of the transmitted beam in the scattering experiment.
This is due to the fact that in a typical scattering experiment it is intensity, or in other words, a square modulus of a complex amplitude, that is measured on the detector.
In Ref. \cite{Hirano1996} it was proposed to measure dynamical phase correction using interferometry measurements based on Bonse-Hart interferometer \cite{Bonse_Hart}.
By recombining two coherent beams passing through empty space and a crystal positioned close to Bragg angle an interference pattern was measured that could be well described in simulation by the presence of the phase of the transmitted beam.
This was still indirect measurement of the phase.
Recently developed coherent x-ray scattering methods, such as ptychography \cite{ptychography1, ptychography2}, can, in principle, provide direct measurements of phase.
Importantly, these methods being highly sensitive to changes of phase, can be applied to small variations of the phase of the transmitted beam.
Such experiment in  which the phase of the transmitted beam in ptychographic measurements was determined was performed recently \cite{Robinson} (see layout of this experiment in Fig. \ref{fig::schematic}).
In this experiment two detectors were used, one in diffraction and one in transmission direction.
Ptychographic measurements were performed both on empty membrane and Au crystalline nanoparticles 100 nm thick at different rocking angles of the sample.
These measurements provided phase information of the transmitted beam that contained two components: the main one due to conventional refraction that does not depend on rocking angle and another one that is much weaker and had an angular dependence on the rocking angle.

Though physical principles of generation of the transmitted wave phase are well understood and can be well simulated using dynamical theory \cite{Authier} their solutions can be quite complicated in some special cases and often do not provide analytical results.
On the contrary, the kinematical theory based on the assumption of a single scattering of the incident beam on the sample, is easier to understand and interpret.
However, kinematical theory does not describe variations of the phase close to Bragg angle in transmitted beam.
To fill this gap, we developed here quasi-kinematical theory that provides a simple analytical description for the phase of the transmitted beam.

Typically thick crystals, with the thickness larger than the so-called extinction length \cite{Authier}, have to be described by the dynamical theory.
At the same time crystals with the thickness much smaller than this extinction length can be safely described by the kinematical theory \cite{Zachariazen}.
So, extinction length provides the typical crystal size for which multiple scattering effects, such as coupling between the transmitted and diffracted wave, becomes important.
We will show that quasi-kinematical approximation developed here gives correct results for the transmitted phase for crystal thicknesses up to extinction length that is significantly beyond the conventional kinematical theory.

In this paper we first revise the concept of the phase of the transmitted beam for x-rays.
Then using full dynamical theory we show how pure kinematical case can be generalized by including refraction and absorption effects.
We also present general dynamical theory expression for the transmitted phase that is especially transparent in the case of a perfect crystal.
We also present simulations based on dynamical theory of the phase of the transmitted beam for different crystal types, reflection orders and incoming photon energies.
Next, we introduce a quasi-kinematical approximation and obtain analytical solution for the dynamical phase contribution in the transmitted beam.
We finalize our work with the summary and outlook section.

\section{General considerations}


\subsection{Non-periodic media}

We will first consider the case of scattering of x-rays on non-periodic media.
In this case refraction coefficient $n$ of matter for x-rays is given by
\begin{equation}
n = \sqrt{1+\chi_0} \simeq 1 + \chi_0/2 \ ,
\label{dynamical::refraction_coefficient}
\end{equation}
where $\chi_0$ is the zero-th Fourier component of the susceptibility $\chi(\textbf{r}, \omega)$ that is connected with the electron density $\rho$ by known expression \cite{Als-Nielsen} $\chi_0 = -r_e \lambda^2 \rho / \pi$, where  $r_e$ is the classical electron radius and $\lambda$ is the x-rays wavelength.
Here we took into account that for x-rays $\chi_0 \ll 1$.
The x-ray wave passing through a slab of material of thickness $d$ will get an additional phase shift
\begin{equation}
E_{out}(d) = E_{in}(z=0) e^{i\varphi(d)}   \ ,
\label{dynamical::phase}
\end{equation}
where the total phase $\varphi(d)$ accumulated while passing the material is given by
\begin{equation}
\varphi(d) = nkd/\gamma = \left(1 + \chi_0/2\right)kd/\gamma \ .
\label{dynamical::phase_2}
\end{equation}
Here $k=\omega/c$ is the vacuum value of the incidence wavevector, $\omega$ is the frequency of x-rays, $c$ is the speed of light and
$\gamma = \cos(\mathbf{n \cdot k})$
is the direction cosine with $\textbf{n}$ being the inward normal to the entrance surface of the material.
We can see from that expression that the phase due to refraction in non-periodic media is given by a simple expression
\begin{equation}
\varphi_{ref}(d) =  \chi_0 kd/\left(2\gamma\right) \ .
\label{dynamical::refracted_phase}
\end{equation}
This is a conventional phase shift due to refraction proportional to an effective thickness of the material $t=d/\gamma$, well known for electromagnetic waves, the only difference for x-rays is that it is negative since $\chi_0 < 0$.

%
%

\subsection{Periodic media}

For x-ray wave passing through periodic media the situation is similar to non-periodic media in most cases except incident angles close to Bragg angle.
As it follows from the dynamical theory \cite{Authier} at these angles an additional dynamical correction $\delta n$ to refractive index $n$ will appear in expression (\ref{dynamical::refraction_coefficient}).
Contrary to expression \eqref{dynamical::refraction_coefficient} it will be depended on the rocking angle $\Delta\theta$.
Asymptotically, far from exact Bragg condition, this phase correction can be expressed as \cite{Authier}
\footnote{Note the difference by a factor of two between this expression and the one given in the book \cite{Authier}.}
\begin{equation}
\delta n \approx - \frac{C^2\chi_h \chi_{\overline{h}}}{4\gamma_0\left( \Delta\theta - \Delta\theta_r\right)\sin2\theta_B} \ .
\label{dynamical::dynamical_phase_asymptotics}
\end{equation}

Here $\chi_h$ and $\chi_{\overline{h}}$ are the Fourier components of the susceptibility of the $h$ and $\overline{h}$ reflections, respectively.
They are connected with the Fourier components of the structure factor $F_h$ and $F_{\overline{h}}$ by well known relations
\begin{equation}
\chi_h = -\Gamma F_h \ ,
\chi_{\overline{h}} = -\Gamma F_{\overline{h}} \
\label{dynamical::chi_h}
\end{equation}
and parameter $\Gamma$ is given by $\Gamma = r_e \lambda^2/\pi V$, where $V$ is the volume of the unit cell.
In expression (\ref{dynamical::dynamical_phase_asymptotics}) $C$ is the polarization coefficient equal to $C=1$ for $\sigma-$ polarization and $C=cos2\theta_B$ for $\pi-$ polarization, $\Delta\theta = \theta - \theta_B$ is the angular deviation from the exact Bragg conditions, $\Delta\theta_r$ is the angular correction due to refraction and $\theta_B$ is the Bragg angle.
We note that this expression is valid for both Bragg and Laue geometry.
As was mentioned above the range of validity of this expression is given by inequality
\begin{equation}
\left| \Delta\theta \right| \gg \left|\chi_h\right| / \sin2\theta_B \ .
\label{dynamical::asymptotic_conditions}
\end{equation}

As it follows from expression (\ref{dynamical::dynamical_phase_asymptotics}) the dynamical phase correction due to the coupling of transmitted and diffracted waves is proportional to the product of Fourier components of susceptibilities $\chi_h \chi_{\overline{h}}$ or structure factors $F_h F_{\overline{h}}$.
It asymptotically decreases as $1/\Delta\theta$ with the increase of the angular deviation from the Bragg angle and is proportional to the first power of a crystal thickness.
Unfortunately, there is no simple relationship describing the phase of transmitted beam for a more general case of an arbitrary thick crystal for the whole angular range near the Bragg angle, or a crystal with deformation field that could also change the values of the transmitted phase.
In the following we will analyze in detail the case of quasi-kinematical approximation when the phase due to dynamical scattering will be considered as a small perturbation to the phase $\varphi_{ref}(d)$ \eqref{dynamical::refracted_phase} due to refraction.
Obtained analytical solutions will be compared with full dynamical simulations and the range of validity of quasi-kinematical approximation will be determined.

\section{Dynamical theory approach}

\subsection{Theory. General equations}

\begin{figure}
	\includegraphics[width=\linewidth]{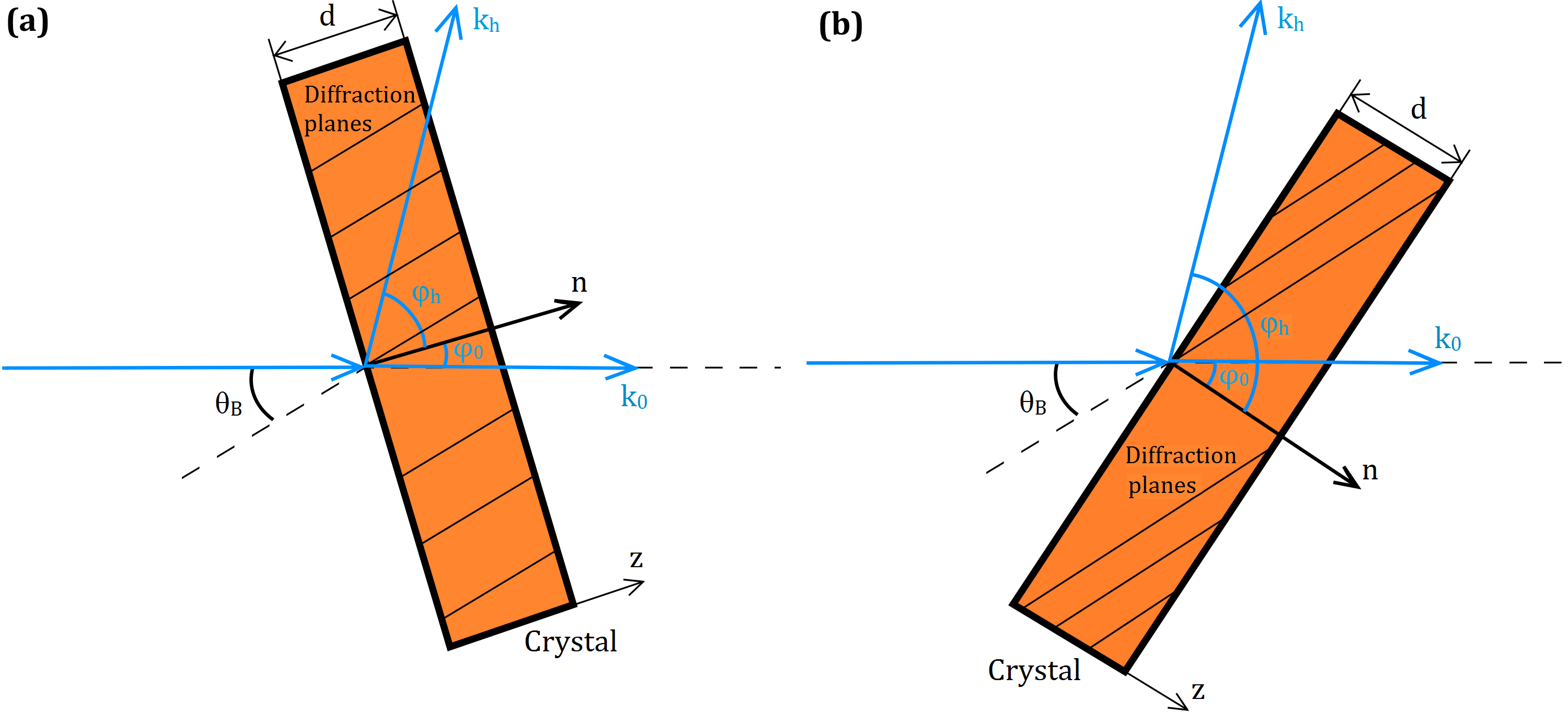}
	\caption{
    \label{fig::scheme}
		(a) Diffraction scattering experiment in Laue geometry on a single crystal of the thickness $d$.
        Here $\theta_B$ is the Bragg angle, $\varphi_0$ and $\varphi_h$ are the angles between the normal $\mathbf{n}$ to the crystal entrance surface, transmitted $\mathbf{k_0}$ and diffracted $\mathbf{k_h}$ wavevectors, respectively.
        (b) Diffraction scattering experiment in Bragg geometry on a single crystal of the thickness $d$.
	}
	\newpage
\end{figure}

In the following we will consider a plane x-ray wave with a wavevector $\mathbf{k}$ incident on a single crystal plate of thickness $d$.
For generality we will consider both Bragg and Laue diffraction geometries (see Fig. \ref{fig::scheme}).
%
%
%


In the dynamical theory, in the two wave approximation \cite{Authier} the wavefield inside the crystal can be presented as a coherent superposition of the transmitted $E_{0s}(\mathbf{r})$ and diffracted $E_{hs}(\mathbf{r})$ waves
\begin{equation}
\label{dynamical::decomposition}
\begin{gathered}
\mathbf{E}(\mathbf{r}) = \sum_s \left[ \mathbf{e}_{0s} E_{0s}(z)e^{i\mathbf{k_0r}} + \mathbf{e}_{hs} E_{hs}(z)e^{i\mathbf{k_hr}} \right] \ ,
\end{gathered}
\end{equation}
where $\mathbf{e}_{0}$ and $\mathbf{e}_{h}$ are polarization unit vectors and $s$ is the polarization index,
$\mathbf{k_0}$ is the incident wavevector and $\mathbf{k_h} = \mathbf{k_0} + \mathbf{h}$ is the diffracted wavevector with $\mathbf{h}=2\pi/\mathbf{H}$ and $\mathbf{H}$ being reciprocal space vector.
Here we also assume that slowly varying amplitudes $E_{0s}(z)$ and $E_{hs}(z)$ have only $z-$dependence.

Propagation of transmitted and diffracted amplitudes in a weakly deformed crystal can be described by the Takagi-Taupin (T-T) equations \cite{Takagi,Taupin,Authier,VartKov}
\begin{subequations}
	\label{dynamical::Takagi}
	\begin{gather}
	\frac{dE_{0s}}{dz} = \frac{ik}{2\gamma_0}\left[ \chi_0 E_{0s}(z) + \chi_{\overline{h}} C e^{i\mathbf{hu}(z)-W(z)} E_{hs}(z) \right] \ ,
	\label{dynamical::Takagi_1} \\
	\frac{dE_{hs}}{dz} = \frac{ik}{2\gamma_h}\left[ (\chi_0 - \alpha) E_{hs}(z) + \chi_{h} C e^{-i\mathbf{hu}(z)-W(z)} E_{0s}(z) \right] \ .
	\label{dynamical::Takagi_2}
	\end{gather}
\end{subequations}
Here $\mathbf{u}(z)$ and $W(z)$ are the strain field and static Debye-Waller factor
(for a perfect crystal $\mathbf{u}(z) = W(z) = 0$),
$\gamma_{0,h} = \cos(\mathbf{n \cdot k}_{0,h})$ are the direction cosines; $\mathbf{n}$ is the inward normal to the entrance surface of the crystal (see Fig. \ref{fig::scheme}).
For Bragg geometry of diffraction, $\gamma_0 > 0$ and $\gamma_h < 0$, and for Laue diffraction $\gamma_0 > 0$ and $\gamma_h > 0$.
Fourier components of the susceptibility are in general complex valued numbers $\chi_{h} = \chi_{hr} +i \chi_{hi}$.
The parameter $\alpha$ characterizes the deviation of the incident wavevector $\mathbf{k_0}$ from the Bragg condition
$\alpha = (k^2_h-k^2_0)/k^2_0 \approx -2\sin{2\theta_B}\Delta\theta$.

The boundary conditions in Bragg geometry are given by
\begin{equation}
E_{0s}(0) = E_s^{in} \ , \
E_{hs}(d) = 0
\label{dynamical::boundary_Bragg}
\end{equation}
and in Laue geometry
\begin{equation}
E_{0s}(0) = E_s^{in} \ , \
E_{hs}(0) = 0 \ .
\label{dynamical::boundary_Laue}
\end{equation}
We will assume in the following a unit amplitude for the incoming beam $E_s^{in}=1$.

The intensity of the diffracted beam, or reflectivity, is defined in Bragg geometry as
\begin{equation}
\label{dynamical::reflectivity_Bragg}
p^B_R(\Delta\theta) = \frac{\gamma_h}{\gamma_0} \left| \frac{E_h(0, \Delta\theta)}{E_0(0, \Delta\theta)} \right|^2 \ .
\end{equation}
and in Laue geometry as
\begin{equation}
\label{dynamical::reflectivity_Laue}
p^L_R(\Delta\theta) = \frac{\gamma_h}{\gamma_0} \left| E_h(d, \Delta\theta) \right|^2 \ .
\end{equation}
%


\subsection{Kinematical solution}

The kinematical solution
\footnote{We should note that strictly speaking in the case of kinematical scattering the transmitted amplitude is equal to unity. Here we include in the transmitted amplitude refraction and normal absorption but neglect dynamical effects of multiple scattering.}
for the transmitted wave $E_{0}(z)$ can be obtained from Eq. \eqref{dynamical::Takagi_1} by neglecting the coupling term with the diffracted amplitude $E_h(z)$.
In this case we obtain from Eq. \eqref{dynamical::Takagi_1}
\begin{equation}
\frac{dE_{0s}}{dz} = i \delta_0  E_{0s}(z) \ ,\text{where} \  \delta_0 = \frac{k \chi_0}{2\gamma_0} \ .
\label{dynamical::kinematical_transm_beam_equation}
\end{equation}
Solution of this equation gives for the x-ray wave on the exit surface of the crystal

\begin{equation}
\label{dynamical::kinematical_transm_beam}
E^{out}_{0s}(d) = e^{i \delta_0 d }  = \exp \left[ i \varphi_{ref}(d)-\mu_0d/2\gamma_0 \right] \ ,
\end{equation}
where
\begin{equation}
\label{dynamical::kinematical_phase}
\varphi_{ref}(d) = Re\left[\delta_0 d \right] = \frac{k\chi_{0r} }{2\gamma_0} d \ ,
\end{equation}
is the phase due to refraction (compare it with expression \eqref{dynamical::refracted_phase}) and
$\mu_0 =k\chi_{0i}$ is the linear absorption coefficient.

Notice that transmitted wave $E_0(z)$ \eqref{dynamical::kinematical_transm_beam} determined in this way does not depend on the deviation angle from the exact Bragg condition $\Delta\theta$.
It is only attenuated by absorption due to imaginary part of the susceptibility $\chi_{0i}$ and has a constant phase shift $\varphi_{ref}(d)$ due to real part of the susceptibility $\chi_{0r}$.
By this treatment we already go beyond the conventional kinematical theory that typically neglects these effects.
However, we still neglected multiple scattering or dynamical effects that will be taken into account below.






\subsection{Dynamical solution}

An expression for the transmitted wave $E^{out}_{0}(d)$ on the exit surface of the crystal in the case of dynamical diffraction can be obtained as a formal solution of the T-T equations (\ref{dynamical::Takagi}) in the following form (see for details \cite{VartKov})
\begin{equation}
\label{dynamical::transmitted wave}
E^{out}_{0}(d,\Delta\theta) = \exp\left[ i \varphi_{ref}-\mu_0d/2\gamma_0 + i \varphi_{dyn}(d,\Delta\theta) \right]  \ ,
\end{equation}
where $\varphi_{dyn}(d,\Delta\theta)$ is the phase contribution due to dynamical scattering given by
\begin{equation}
\label{dynamical::phase_dynamical}
\varphi_{dyn}(d,\Delta\theta) =  - \frac{1}{L_{ex}} Re \left[ \int_0^d dz^{'} C_1 R(z^{'}, \Delta\theta) \right]  \ .
\end{equation}
Here $R(z, \Delta\theta)$ is the scattering amplitude (see for details Appendix A) defined as \cite{VartKov}
\begin{equation}
\label{dynamical::scattering_amplitude_definition}
R(z,\Delta\theta) = \frac{1}{\sqrt{\beta}Y } \left(  \frac{E_h(z,\Delta\theta)}{E_0(z,\Delta\theta)}\right) e^{i\mathbf{hu}(z)} \ ,
\end{equation}
In Eqs. (\ref{dynamical::phase_dynamical}) and (\ref{dynamical::scattering_amplitude_definition}) parameter
$\beta=\gamma_0/|\gamma_h|$  for Bragg and
$\beta=\gamma_0/\gamma_h$ for Laue geometries,
$C_1 = C (1-ip)exp{[-W(z)]}$ with
$p=-X_i/X_r$, and parameter
$Y = \sqrt{\chi_h/\chi_{\overline{h}}} = |Y|exp(i\Phi_Y)$.
For a centrosymmetric crystal with a monoatomic lattice $|Y|=1$, $\Phi_Y=0$.
The following parameters have been also introduced:
$X_r = Re\sqrt{\chi_h\chi_{\overline{h}}}$ and $X_i = Im\sqrt{\chi_h\chi_{\overline{h}}}$.
Extinction length in Eq. (\ref{dynamical::phase_dynamical}) is defined as
\footnote{It should be noted that our choice of extinction length differs from that commonly used \cite{Authier} by the factor $\pi$.}
\begin{equation}
\label{dynamical::extinction_length}
L_{ex} = \frac{\lambda \gamma_0}{\pi \sqrt{\beta} X_r}  \ .
\end{equation}

In the case of a perfect thick crystal $R(z,\Delta\theta) = R_0(\Delta\theta)$ and does not depend on the crystal thickness (see for details Appendix A)
and we obtain for the dynamical phase (\ref{dynamical::phase_dynamical})
\begin{equation}
\label{dynamical::phase_dynamical_perfect_crystal}
\varphi_{dyn}(d,\Delta\theta) =  - \frac{d}{L_{ex}} Re \left[ C_1 R_0(\Delta\theta) \right]  \ .
\end{equation}
Below we will present results of dynamical simulations using crystals of different thickness, reflection order and incident photon energy based on that approach.


\subsection{Simulations}

We performed full dynamical simulations of the intensity of the diffracted beam and phase of the transmitted beam for gold and silicon crystals of different thickness in Laue geometry.
The diffraction scheme was considered asymmetrical in both cases with the incident beam perpendicular to the entrance surface of the crystal as shown in Fig. \ref{fig::schematic} that corresponds to the choice of angles $\varphi_0 = 0$ and $\varphi_h = 2\theta_B$ in Fig. \ref{fig::scheme}.
For these simulations Fourier components of susceptibilities were obtained from Ref. \cite{x0h} and all scattering parameters are summarized in Table I.
We want to note here that for scattering conditions considered here the extinction length for Si 111 crystal $L^{Si}_{ex}=6.1$ $\mu$m was about an order of magnitude larger than for Au 111 crystal $L^{Au}_{ex}=610$ nm.

The reflectivity curves and corresponding phases of the transmitted beam as a function of the rocking angle $\Delta\theta$ were simulated using dynamical theory approach.
Results of these simulations for Au and Si crystals of different thickness from $d=0.2 \ L_{ex}$ to $d=1.5 \ L_{ex}$ are presented in Figs.  \ref{fig::ampl_phase_thin} and \ref{fig::ampl_phase_thick}.
As it follows from these simulations the angular variation as well as the magnitude of the transmitted wave phase are similar for both crystals for the same ratio of $d/L_{ex}$.
Higher is the value of this ratio stronger is the reflectivity curve and values of the phase modulation.
The only difference is the angular range in which these variations of phase are significant.
It is much broader in the case of Au crystal and is quite narrow in the case of Si crystal, that can be explained by the difference in real part of the susceptibility (see Table 1.)
In addition, due to difference in extinction length, the actual thickness of each crystal (Au or Si) is significantly different (see Table I).

Simulations performed in the frame of dynamical theory for different reflection orders and incoming photon energies for Au crystals of thickness $d=300$ nm are presented in Appendix A.


There are also some other common features that can be observed in these simulations.
The maximum of the reflectivity curve as well as the angular position of the sign change in the phase are shifted from the exact Bragg position ($\Delta\theta=0$) to positive values by
\begin{equation}
\theta_{ref}=\mp \frac{\chi_{0r}(1\pm\beta)}{2\beta\sin2\theta_B} \ ,
\label{refraction_angle}
\end{equation}
where as before the upper sign corresponds to Bragg diffraction and the lower one to Laue diffraction.
This shift due to refraction is well known in dynamical theory \cite{Authier}.
We want to note that in the case of symmetrical Laue diffraction, when parameter $\beta=1$, this shift due to refraction is  zero.

One important feature that we can observe in our simulations is that on the left side of the rocking curve the phase is positive and goes through its maximum and on the right side of the rocking curve it has an opposite behavior.
In addition, we can observe that for a thick crystal with $d=1.5 \ L_{ex}$ negative modulation of phase is slightly lower than the same positive valued modulation.
To understand the physical reasons of such behavior we should recall that according to the dynamical theory \cite{Authier} in the vicinity of the Bragg angle in Laue geometry two standing waves are generated.
One, weakly absorbing, with its nodes at atomic planes and another one, strongly absorbing, with its antinodes at atomic planes.
In thin crystals both waves contribute with the same strength, however, in a thick crystal the first wave starts to dominate.
Our simulations (see Fig. \ref{fig::weakly_strongly_abs_waves}) have shown that on the left side of the rocking curve the wavefield is dominated by the weakly absorbed wave and on the right side by the strongly absorbed wave.
This also explains the fact that we have on the left side of the rocking curve positive relative values of phase (that means that the wave is accelerated) and we have negative relative values (that means that the wave is retarded) on the right side.
This is due to the fact that when the standing wave is with its nodes on the atomic planes, than transmitted wave probes an effectively lower electron density than an averaged electron density.
And on the contrary, when the standing wave is with its antinodes on the atomic planes the transmitted wave probes an effectively higher electron density than an average electron density.

As it was mentioned before these results were obtained using full dynamical treatment.
At the same time we can easily observe that in the case of thin crystals with $d/L_{ex}<1$ the phase variations of the transmitted beam have similar behavior and are only scaled with the value of the ratio $d/L_{ex}<1$.
It will be very useful if a simple analytical expression can be derived that could explain the behavior of this phase variation for thin crystals.
In the next section we show how such expression can be obtained from the T-T equations \eqref{dynamical::Takagi}.

\section{Quasi-kinematical approximation}

We will consider now a thin crystal in Bragg, or Laue geometry, such that the transmitted wave does not differ significantly from the incident wave.
To characterize the difference between kinematical and dynamical solutions for the transmitted wave, a small parameter
\begin{equation}
|\delta_{dyn}(z,\Delta\theta)|<<1
\label{approx::approximation}
\end{equation}
can be introduced as
\footnote{Compare this expression with the one for the transmitted beam in kinematical approximation \eqref{dynamical::kinematical_transm_beam}.}
\begin{equation}
E_{0}(z, \theta) \approx exp\left[ i \delta_0 d + i\delta_{dyn}(z,\Delta\theta) \right] \ .
\label{approx::assumption}
\end{equation}
where parameter $\delta_0$ is defined in Eq. \eqref{dynamical::kinematical_transm_beam_equation}.
Now additional contribution to the phase due to diffraction is given by the real part of $\delta_{dyn}(z,\Delta\theta)$
\begin{equation}
 \delta\varphi_{dyn}(z,\Delta\theta) = Re\left[\delta_{dyn}(z,\Delta\theta) \right] \ .
\label{approx::dyn_phase_definition}
\end{equation}
The imaginary part of $\delta_{dyn}(z,\Delta\theta)$ will give contribution to an interference absorption coefficient $\mu_{in}(\Delta\theta)$ (see for details \cite{VartKov}).


By substituting Eq. (\ref{approx::assumption}) into T-T equations (\ref{dynamical::Takagi}) (see for details Appendix B), it is possible to obtain the following solution for the dynamical contribution to the transmitted beam in the quasi-kinematical approximation
\begin{subequations}
	\begin{gather}
	\delta_{dyn} (d, \Delta\theta) = - \frac{C^2 \chi_h \chi_{\overline{h}} (kd)^2}{8 \gamma_0 \gamma_h}
                                \frac{1}{\Omega} \left[1 -  e^{i\Omega} \left( \frac{\sin{\Omega }}{\Omega} \right) \right] \ ,
 \\
    \Omega(\Delta\theta) = Q(\Delta\theta)d/2 \ ,
	\end{gather}
\label{approx::phase}
\end{subequations}
where $Q(\Delta\theta) = (2/L_{ex})\left[ y(\Delta\theta) +iy_0 \right]$ is a momentum transfer due to angular deviations from the Bragg angle and dimensionless angular parameters $y(\Delta\theta)$ and $y_0$ are defined in Eqs. (\ref{dynamical::angular_parameter}) and (\ref{dynamical::y_0_y_u}).

Expression \eqref{approx::phase} for the dynamical contribution to the transmitted beam can be also written in a more compact form by introducing extinction length $L_{ex}$ \eqref{dynamical::extinction_length} (see Appendix B)
\begin{equation}
\delta_{dyn} (d, \Delta\theta) =  - \frac{ C_1^2}{2} \left(\frac{d}{L_{ex}}\right)^2
 \frac{1}{\Omega} \left[1 -  e^{i\Omega} \left( \frac{\sin{\Omega }}{\Omega} \right) \right] \ ,
\label{approx_phase_2}
\end{equation}
where parameter $C_1$ is defined after equation \eqref{dynamical::scattering_amplitude_definition}.
Taking into account definition of the phase of the transmitted wave \eqref{approx::dyn_phase_definition} we obtain
\begin{equation}
\delta\varphi_{dyn} (d, \Delta\theta) = - \frac{ C^2}{2} \left(\frac{d}{L_{ex}}\right)^2
  \frac{1}{\Omega} \left\{ \left( 1-p^2 \right)
 \left[1 -  \cos{\Omega} \left( \frac{\sin{\Omega }}{\Omega} \right) \right]
 - 2p \sin{\Omega} \left( \frac{\sin{\Omega}}{\Omega} \right) \right\} \ .
\label{approx_phase_3}
\end{equation}

As it follows from these results dynamical contribution to the refraction coefficient is proportional to the product $\chi_h \chi_{\overline{h}}$ that is similar to Eq. \eqref{dynamical::dynamical_phase_asymptotics}.
Another important feature is that in quasi-kinematical approximation the dynamical phase contribution is proportional to the square of the ratio $d/L_{ex}$ (see Eq. \eqref{approx_phase_3}).

In the frame of the same approximations it is possible to obtain expression for the amplitude of the diffracted wave (see for details Appendix B)
\begin{equation}
E_h(d,\Delta\theta) = i E_h^0  e^{i \delta_h z} e^{-i\Omega} \left( \frac {\sin\Omega}{\Omega} \right) \ ,
\label{approx::diffr_wave}
\end{equation}
where $E_h^0 = Ckd\chi_{h}/(2\gamma_h)$.
Substituting this expression in  Eqs. (\ref{dynamical::reflectivity_Bragg}, \ref{dynamical::reflectivity_Laue}) we obtain well known expression \cite{Zachariazen} for the reflectivity in kinematical approximation in Laue or Bragg geometry in quasi-kinematical approximation
\begin{equation}
p_R(\Delta\theta) = \frac{\gamma_h}{\gamma_0} \left| E_h^0 \right|^2  \frac {\sin^2\Omega}{\Omega^2}  \ .
\label{approx::reflectivity}
\end{equation}

We consider now different asymptotics of the obtained solutions.
The dynamical correction in the case of big angular deviations from the exact Bragg angle can be determined from expression \eqref{approx::phase}.
For these angles we can drop off fast oscillating second term in square brackets in expression \eqref{approx::phase} and after substituting the value of the angular parameter $\Omega(\Delta\theta)$ (we neglect here its imaginary part) we obtain

\begin{equation}
\delta_{dyn} (d, \Delta\theta) \approx - \frac{C^2 \chi_h \chi_{\overline{h}} (kd)}{4 \gamma_0 \sin{2\theta_B} \left( \Delta\theta -\theta_{ref} \right)} \ ,
\label{approx::phase_asymptotics}
\end{equation}
where $\theta_{ref}$ is the angular correction due to refraction \eqref{refraction_angle}.
Comparison of this expression with the one obtained from the full dynamical theory (see Eq. \eqref{dynamical::dynamical_phase_asymptotics}) shows that they completely coincide.

\begin{figure}
	\includegraphics[width=\linewidth]{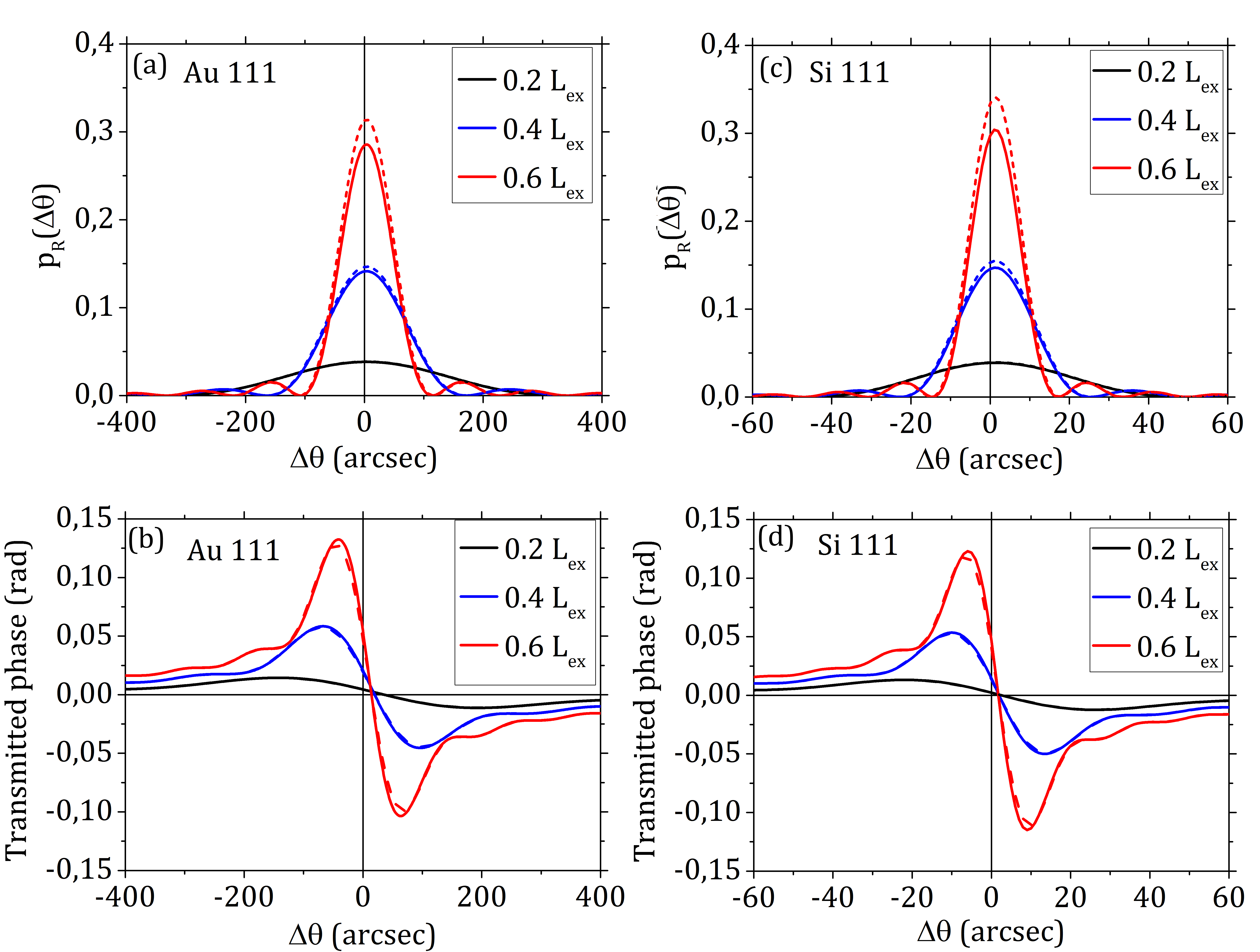}
	\caption{\label{fig::ampl_phase_thin}
		Reflectivity $p_R(\Delta\theta)$ (a,c) and the phase $\varphi_{dyn}(\Delta\theta)$ (b,d) of the transmitted beam in
        Laue geometry as a function of the rocking angle $\Delta\theta = \theta - \theta_B$.
        Simulations were performed for Au 111 and Si 111 crystals of different thickness $d=0.2\ L_{ex}$, $0.4\ L_{ex}$, and $0.6\ L_{ex}$.
        Full lines dynamical theory simulations, dashed lines simulations performed in the frame of quasi-kinematical approximation.
        Parameters of simulations are listed in Table I.
     	}
	\newpage
\end{figure}

We can also determine behavior of the dynamical contribution to the transmitted beam at small values of the angular parameter $\Omega(\Delta\theta)$.
From expression \eqref{approx::reflectivity} it follows that reflectivity of the diffracted wave has its maximum at $\Omega(\Delta\theta)=0$.
In the limit of $\Omega(\Delta\theta) \rightarrow 0$ we obtain for $\delta_{dyn} (d, \Delta\theta)$ in Eq. \eqref{approx_phase_2}
\begin{equation}
\delta_{dyn} (d, \Delta\theta) \rightarrow  - \frac{ C_1^2}{2} \left(\frac{d}{L_{ex}}\right)^2
 \left[2/3 \Omega - i \right] \ .
\label{approx_phase_small_angle}
\end{equation}
Substituting the values of parameters $C_1$ and $\Omega(\Delta\theta)$ we obtain for the phase of the transmitted beam
%
\begin{equation}
\delta\varphi_{dyn} (d, \Delta\theta)  \rightarrow  - \frac{ C^2}{3} \left(\frac{d}{L_{ex}}\right)^3
\left\{ \left( 1-p^2 \right) y(\Delta\theta) + 2p \left[ y_0 - \frac{3}{2}\left(\frac{L_{ex}}{d}\right) \right] \right\} \ .
 \label{approx_phase_small_angle_2}
\end{equation}
Taking into account that $y(\Delta\theta) \sim \Delta\theta$ we see that it is exactly the angular dependence that we observed at small deviations of angular parameter while performing dynamical simulations for thin crystals as shown in Fig. \ref{fig::ampl_phase_thin}.

Direct comparison of expressions (\ref{approx::phase}) and (\ref{approx::diffr_wave}) shows that the dynamical contribution $\delta_{dyn} (d, \Delta\theta)$  in quasi-kinematical approximation can be expressed through the amplitude of the diffracted wave as
\begin{equation}
\delta_{dyn} (d, \Delta\theta) =  - \frac{C_1^2}{2} \left(\frac{d}{L_{ex}}\right)^2
 \frac{1}{\Omega} \left[1 + i \eta(\Delta\theta)  E_h(d,\Delta\theta) \right] \ ,
\label{approx::phase_diffr_ampl}
\end{equation}
where the following angular parameter is introduced
\begin{equation}
\eta (\Delta\theta) = \frac{2\gamma_h}{C (kd) \chi_h}  e^{-i \delta_h d } e^{2i\Omega}\ .
\label{approx::phase_diffr_ampl_2}
\end{equation}

As a result of our analysis (see Eqs. (\ref{approx::phase}), (\ref{approx::phase_diffr_ampl}) and Appendix C) we see that the phase $\delta\varphi_{dyn} (d, \Delta\theta)$ modulations grow as a second power of a ratio of a crystal thickness to extinction length $L_{ex}$.
At the same time quasi-kinematical approximation is valid if condition (\ref{approx::approximation}) is satisfied that gives for the  maximum crystal thickness $d_{max}$
\begin{equation}
z << d_{max}=\frac{\sqrt{2}}{|C_1|}L_{ex} \ .
\label{approx::thin}
\end{equation}
It follows from Eq. (\ref{approx::thin}) that larger is the extinction length for thicker crystals quasi-kinematical approximation is valid.
Such conditions can be obtained by using higher order reflections and shorter wavelengths.

%
%

In our simulations, we indirectly used the fact that the crystal lateral size $L$ is large enough to neglect boundary effects.
More specifically, it should be larger than the base of the Bormann fan \cite{Authier}
\begin{equation}
L > d\frac{\sin{2\theta_B}}{\gamma_0\gamma_h} \ .
\label{approx::large}
\end{equation}
For example, for Au crystal of 100 nm thickness used in experiment \cite{Robinson} this condition means that the lateral size of a crystal should be larger than $70$ nm.
That condition was safely satisfied in this experiment where lateral size of the crystal was $250$ nm.

\begin{figure}
	\includegraphics[width=\linewidth]{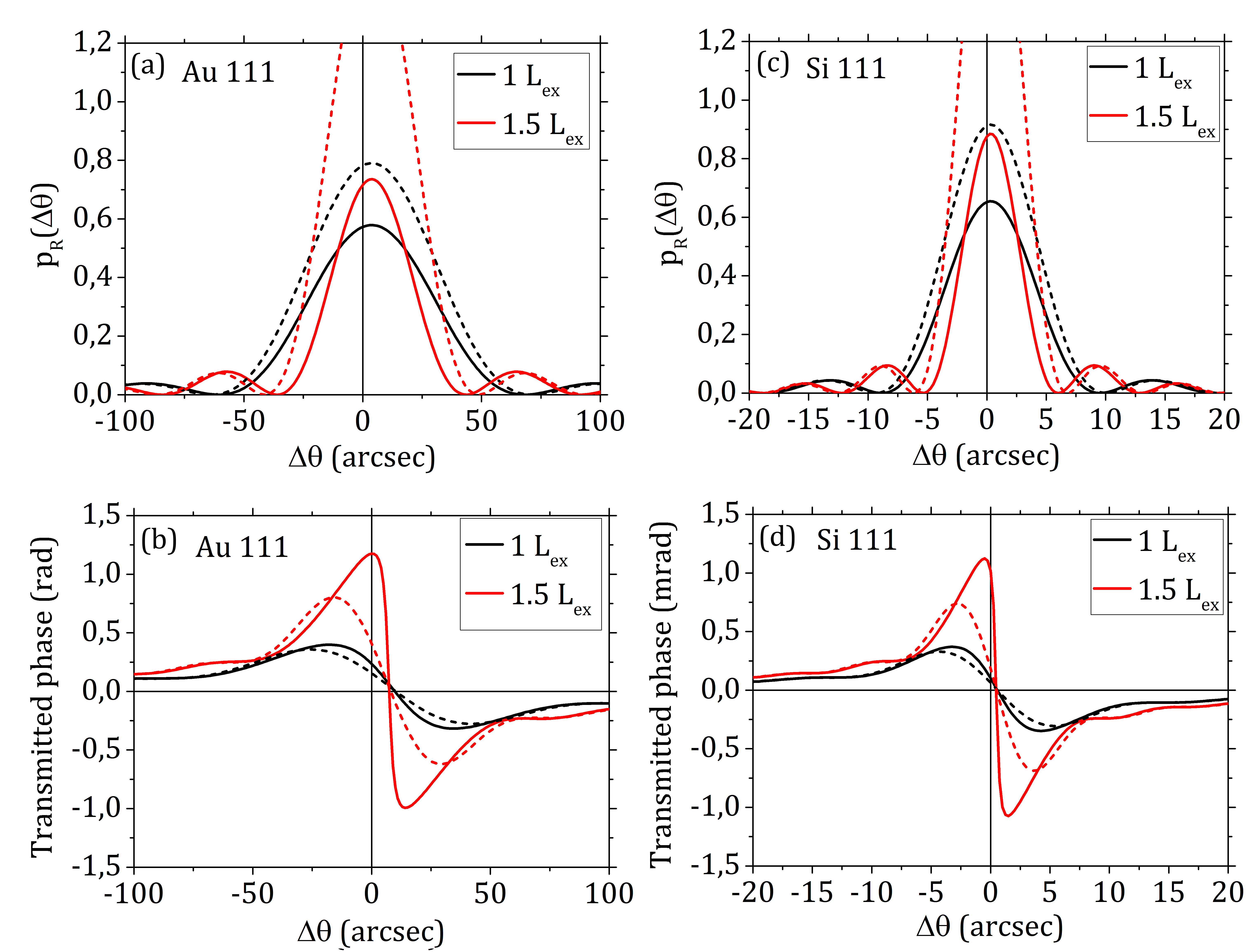}
	\caption{\label{fig::ampl_phase_thick}
		Same as in Fig. \ref{fig::ampl_phase_thin} for crystal thickness $d= L_{ex}$ and $1.5\ L_{ex}$.
	}
	\newpage
\end{figure}

\begin{figure}
	\includegraphics[width=\linewidth]{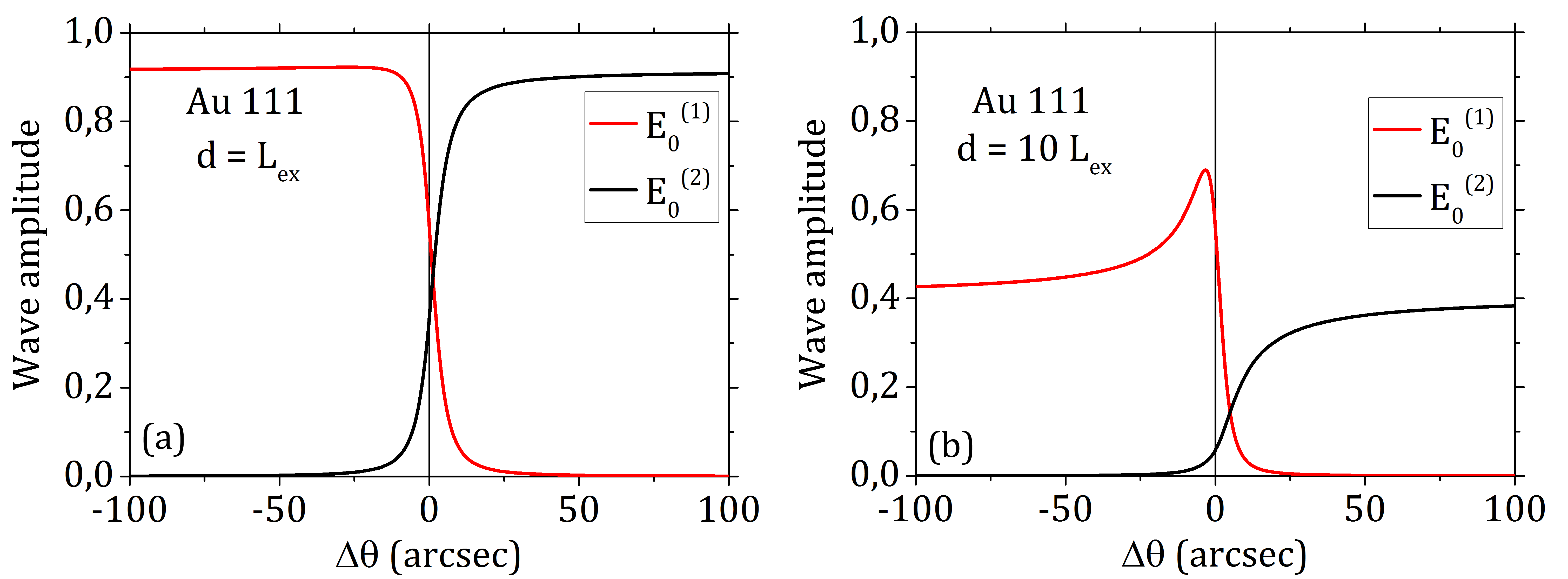}
	\caption{
		Angular dependance of the amplitude of the weakly absorbing wave $E_0^{(1)}$ (red curve) and strongly absorbing wave
        $E_0^{(2)}$ (black curve) for different crystal thickness $d=L_{ex}$ (a) and $d=10 \ L_{ex}$ (b).
        In both cases simulations were performed for Au 111 reflection and photon energy 8.5 keV.
        It is well seen the difference in the amplitudes of the waves for the thick crystal.
    \label{fig::weakly_strongly_abs_waves}
	}
	\newpage
\end{figure}

\begin{figure}
	\includegraphics[width=\linewidth]{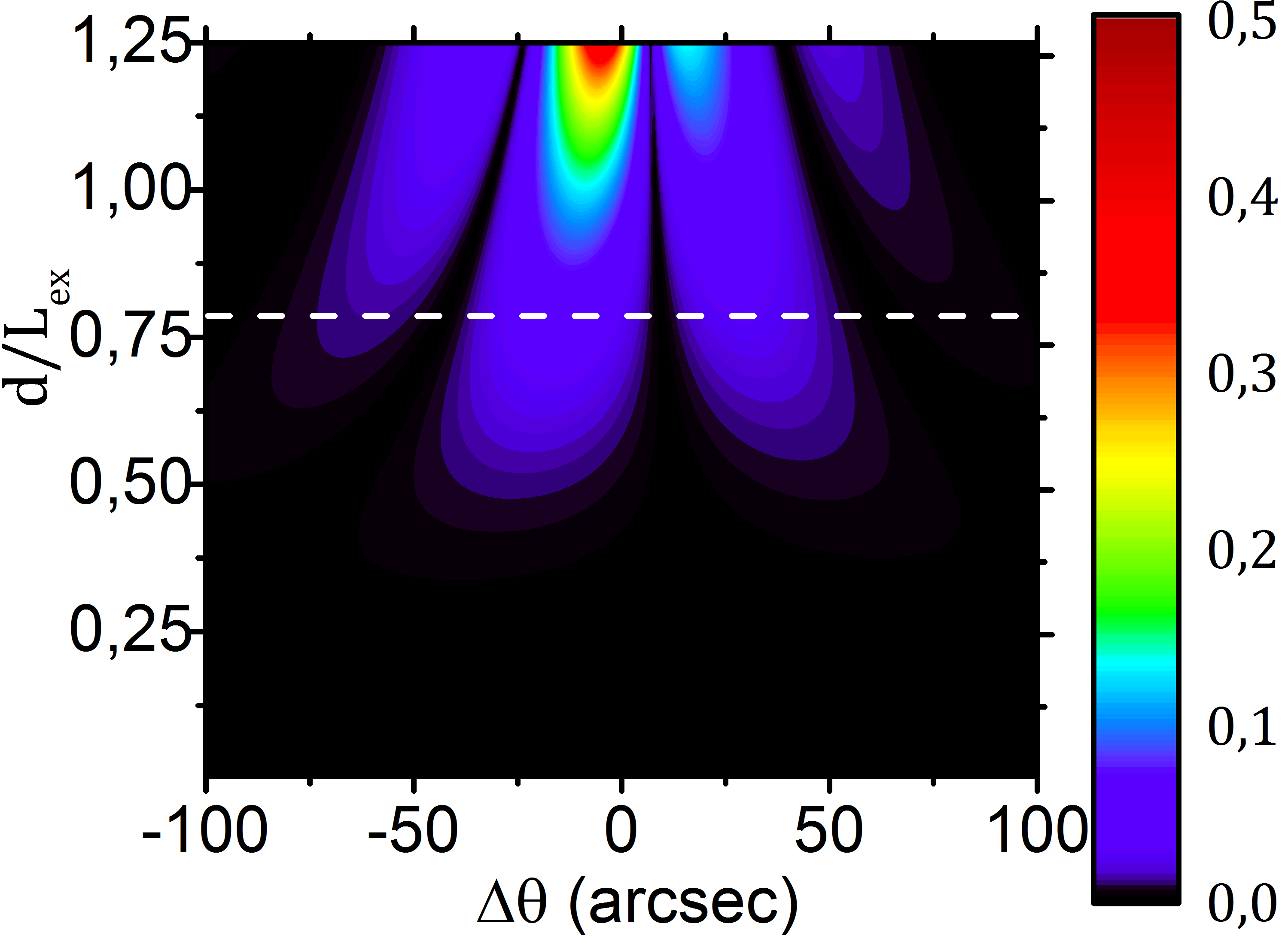}
	\caption{\label{fig::kin_dyn}
		Comparison between the dynamical theory and quasi-kinematical approximation.
		(a) The phase of the transmitted wave simulated according to the dynamical theory (black curve) and quasi-kinematical
        approximation (red curve) for Au 111 reflection, 8.5 keV incident photon energy, and crystal thickness $d = 1.5\  L_{ex}$.
		(b) Relative error $\varepsilon$ in the phase defined in Eq. \eqref{approx::error} calculated using a
        quasi-kinematical approximation as compared to the dynamical theory as a function of rocking angle $\Delta\theta$ and relative crystal thickness $d/L_{ex}$.
        Dashed horizontal line correspond to 5\% difference in the two phases that appears at crystal thickness $d \simeq 0.8\ L_{ex}$.
	}
	\newpage
\end{figure}

We compared the obtained quasi-kinematical result with the exact dynamical solution discussed in the previous section.
The dynamical phase correction in quasi-kinematical approximation given by Eq. (\ref{approx_phase_3}) was calculated for the same set of parameters as in simulations presented in Fig. \ref{fig::ampl_phase_thin} and Fig. \ref{fig::ampl_phase_thick}.
Results of these simulations are shown in these figures by dashed lines.
Our results showed that the difference between quasi-kinematical and dynamical case was less than 3\% for crystal thicknesses up to $0.6 L_{ex}$ (see Fig. \ref{fig::ampl_phase_thin}) and it becomes significant for crystal thicknesses above extinction length (see Fig. \ref{fig::ampl_phase_thick}).
It is interesting to note that even for crystals with the thickness $d=L_{ex}$ deviation of the exact dynamical simulation and simulation performed in the frame of quasi-kinematical approximation is not so strong, though diffracted curves already differ substantially (see Fig. \ref{fig::ampl_phase_thick}).
To determine the range of parameters were the quasi-kinematical approach can be safely used we performed simulations for Au 111 crystal with the thickness varying from zero to $d=1.25 \ L_{ex}$.
We defined an error function $\varepsilon$ between two types of simulations as
\begin{equation}
\varepsilon = \left| \frac{\delta\varphi_{dyn}(d,\Delta\theta) - \varphi_{dyn}(d,\Delta\theta)}{\varphi_{dyn}(d,\Delta\theta)} \right| \ ,
\label{approx::error}
\end{equation}
where $\delta\varphi_{dyn}(d,\Delta\theta)$ is the dynamical phase contribution in quasi-kinematical approximation and $\varphi_{dyn}(d,\Delta\theta)$ is the same phase simulated with the full dynamical theory.
Results of these simulations are presented in Fig. \ref{fig::kin_dyn}.
As it follows from these simulations quasi-kinematical approach can be safely used with an error less than 5\%  up to Au crystal thicknesses $d \simeq 0.8 \ L_{ex}$.
Similar results were obtained also for Si crystal.

\section{Summary}

In summary, the phase variation in the transmitted beam close to Bragg conditions was analyzed for Au and Si crystals of different thickness, reflection and incoming photon energy using dynamical theory.
%
%
It was demonstrated that in the frame of kinematical theory it was not possible to observe phase variations in the transmitted beam close to Bragg angle.
To perform analysis of scattering in thin crystals quasi-kinematical approximation was introduced.
General analytical solution for the phase of the transmitted beam in the whole range of rocking angles was obtained in this limit.
It was determined that in the frame of this approximation the magnitude of the phase variation depends quadratically on the crystal thickness.
These findings can be used in future ptychographic experiments performed on thin crystals where the phase of the transmitted beam can be determined directly from the analysis of the scattered radiation.

\section{Acknowledgements}

We greatly acknowledge discussions with I.K. Robinson who has pointed out our attention to this problem, we are also thankful to G. Materlik, M. Hart and A. Shabalin for fruitful discussions, we acknowledge support of the project and fruitful discussions with E. Weckert and careful reading of the manuscript by D. Novikov. This work was partially supported by the Virtual Institute VH-VI-403 of the Helmholtz Association.

\bibliography{references}

\newpage

\appendix

\section{Dynamical theory treatment}

\subsection{Scattered amplitude in dynamical theory}

The amplitude $R(z,\Delta\theta)$ (\ref{dynamical::scattering_amplitude_definition}) can be determined in the general case of the deformed crystal as a solution of a Riccati type of equation (see for details \cite{VartKov})
\begin{equation}
\label{dynamical::scattering_amplitude}
\mp i L_{ex} \frac{d R(z,\Delta\theta)}{dz} = 2\left[ -y(\Delta\theta)-iy_0 +y_{u} \right]  R(z,\Delta\theta) + C_1 \left[1 \pm  R^2(z,\Delta\theta)\right] \ ,
\end{equation}
where upper sign corresponds to Bragg diffraction and the lower one to Laue diffraction.
Here the angular deviation from the Bragg position is defined by the dimensionless parameter,
\begin{equation}
\label{dynamical::angular_parameter}
y(\Delta\theta) = \sqrt{\beta} \frac{\sin{2\theta_B} \cdot \Delta\theta}{X_r} \pm \frac{\chi_{0r}(1 \pm \beta)}{2\sqrt{\beta}X_r}  \ ,
\end{equation}
parameters
\begin{equation}
\label{dynamical::y_0_y_u}
y_0 = \pm \frac{\chi_{0i}(1 \pm \beta)}{2\sqrt{\beta}X_r} \ \text{and} \  y_u(z) = \pm  \frac{L_{ex}}{2} \frac{d \left( \mathbf{hu}(z) \right)}{dz}
\end{equation}
define the attenuation of x-rays due to photoelectric absorption and the shift of the Bragg position caused by deformation in the crystal.
Boundary conditions for the amplitude $R(z,\Delta\theta)$ in equation (\ref{dynamical::scattering_amplitude}) are defined now on one surface.
For Bragg geometry $R(d,\Delta\theta) = 0$ and for Laue geometry $R(0,\Delta\theta) = 0$.

In some special cases (for example constant strain) analytical solutions for the amplitude $R(z,\Delta\theta)$ can be obtained (see, for example, Ref. \cite{VartKov}).
In general case of an arbitrary strain field $\mathbf{u}(z)$ this amplitude can be determined only numerically.

%
%

For perfect thick crystal solution of Eq. (\ref{dynamical::scattering_amplitude}) gives \cite{VartKov}
\begin{equation}
R_0(\Delta\theta) = \mp \frac{1}{C_1}\left[\left(-y(\Delta\theta)-iy_0\right) + \sqrt{\left(y(\Delta\theta) + iy_0\right)^2 \mp C^2_1} \right] \ ,
\end{equation}
where the branch with the positive imaginary part is chosen for the square root and as before the upper sign corresponds to Bragg diffraction and the lower one to Laue diffraction.

It is possible to show that in this case of a perfect crystal of arbitrary thickness it is also existing an analytical solution for the dynamical amplitude $R(\Delta\theta)$ (see for details Ref. \cite{VartKov}).
Unfortunately, this solution is complicated for the direct analysis.

\subsection{Phase of the transmitted beam for different reflection orders and incoming photon energies}

\begin{figure}
	\includegraphics[width=\linewidth]{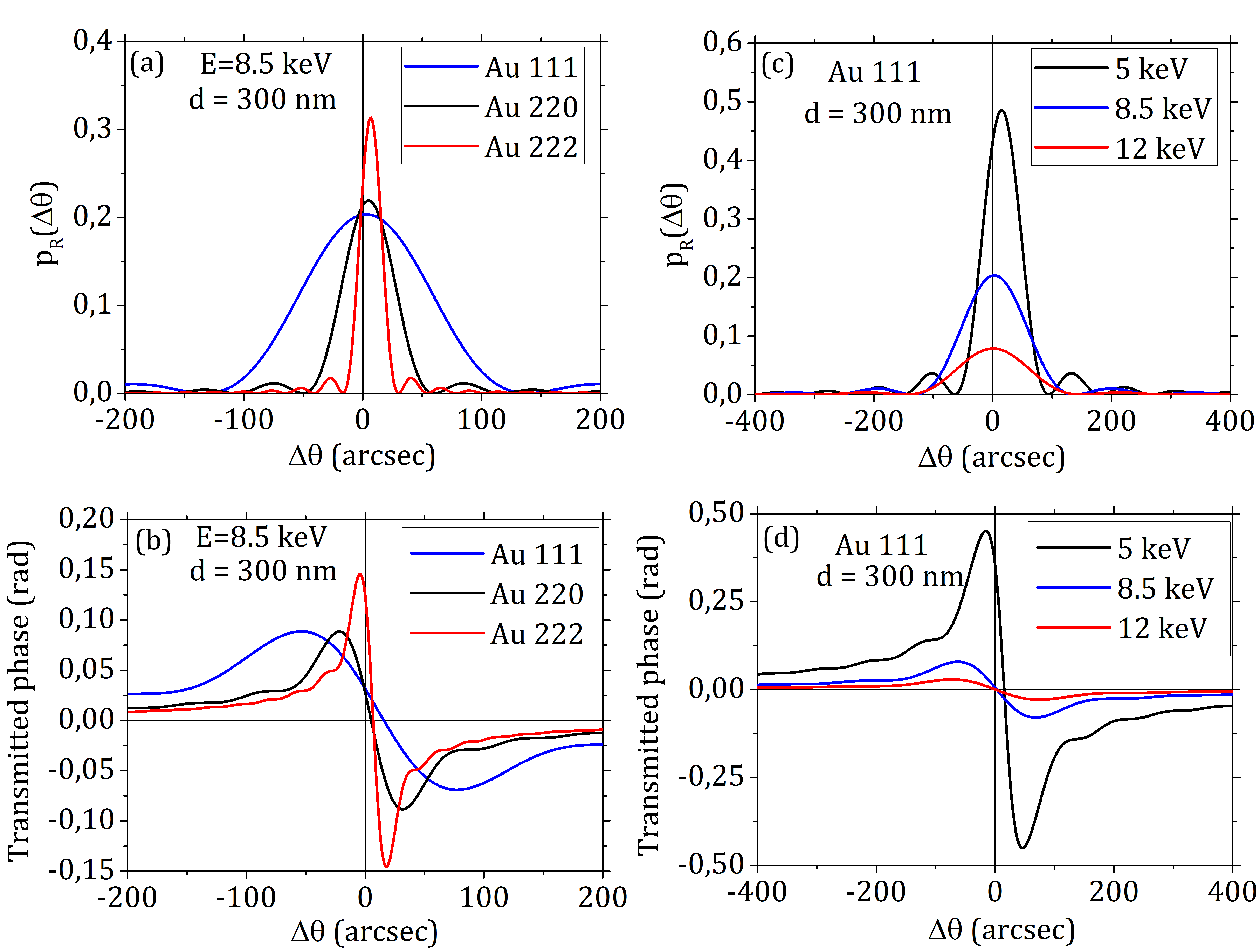}
	\caption{\label{fig::refl_energy}
		Reflectivity $p_R(\Delta\theta)$ (a,c) and the phase $\varphi_{dyn}(\Delta\theta)$ (b,d) of the transmitted beam in
		Laue geometry as a function of the rocking angle $\Delta\theta = \theta - \theta_B$.
		Simulations were performed using dynamical theory approach for a Au crystal of thickness $d=300$ nm.
		(a,b) Results of simulations for the incident photon energy 8.5 keV and different reflection orders 111, 220 and 222 in a Au crystal.
		(c,d) Results of simulations for a Au 111 crystal and different incident photon energies of 5 keV, 8.5 keV, and 12 keV.
		Parameters of simulations are listed in Tables II and III.
	}
	\newpage
\end{figure}

Variations of intensity of the diffracted beam and phase of the transmitted beam for different reflection orders as well as for different incoming photon energies for Au crystal are presented in Fig. \ref{fig::refl_energy}.
We considered here 111, 220, and 222 reflections in Au crystal and incoming photon energies of 5 keV, 8.5 keV and 12 keV.
In the first case the incident photon energy was 8.5 keV for all reflections and in the second Au 111 reflection was considered for all energies.
In all cases the crystal thickness was $d=300$ nm and the diffraction scheme was considered the same as shown in Fig. \ref{fig::scheme}.
All other scattering parameters are listed in Tables II and III.
We can see here that for higher reflections and lower incident photon energies the angular range of the phase variations is becoming more narrow and variations itself become stronger that can be explained by the change of the extinction length (see Table II).

\section{Derivation of the quasi-kinematical approximation}

To derive expression for the dynamical correction $\delta_{dyn}(z,\Delta\theta)$ in transmitted wave $E_0(z, \Delta\theta)$ (\ref{approx::assumption}) in the quasi-kinematical approximation we start with the T-T equations (\ref{dynamical::Takagi}) where we perform the following substitution
\begin{gather}
\label{A1:Transm}
E_0(z) = E^{'}_0(z)e^{i\delta_0 z } \ ,
\text{where} \ \delta_0 = \frac{k\chi_0}{2\gamma_0}
\\
\label{A1:Diffr}
E_h(z) = E^{'}_h(z)e^{i\delta_h z } \ ,
\text{where} \ \delta_h = \frac{k(\chi_0 - \alpha)}{2\gamma_h}
\end{gather}
that leads to the following form of the T-T equations for the amplitudes $E^{'}_0(z)$ and $E^{'}_h(z)$
\begin{subequations}
	\label{dynamical::Takagi_mod}
	\begin{gather}
	\frac{dE^{'}_{0}}{dz} = \left(\frac{ik}{2\gamma_0}\right) C \chi_{\overline{h}} e^{i\textbf{hu}(z)}e^{-W(z)}e^{iQz} E^{'}_h (z) \ ,
	\label{dynamical::Takagi_mod_1} \\
	\frac{dE^{'}_{h}}{dz} = \left(\frac{ik}{2\gamma_h} \right) C \chi_{h} e^{-i\textbf{hu}(z)}e^{-W(z)}e^{-iQz}  E^{'}_0 (z) \ .
	\label{dynamical::Takagi__mod_2}
	\end{gather}
\end{subequations}
Here parameter $Q(\Delta\theta) = (2/L_{ex})\left[ y(\Delta\theta) +iy_0 \right]$ is a momentum transfer due to an angular deviation $\Delta\theta$ from the Bragg angle and dimensionless angular parameters $y(\Delta\theta)$ and $y_0$ are defined in Eqs. (\ref{dynamical::angular_parameter}) and (\ref{dynamical::y_0_y_u}).

Now we consider that in the quasi-kinematical approximation the transmitted wave $E^{'}_0 (z, \Delta\theta)$ can be presented as
\begin{equation}
E^{'}_0 (z, \Delta\theta) = e^{i \delta_{dyn}(z, \Delta\theta)} \ ,
\label{...}
\end{equation}
where the dynamical correction satisfies the condition $\left| \delta_{dyn}(z ,\Delta\theta) \right| \ll 1$.
Substituting this expression for the transmitted wave into the T-T equations (\ref{dynamical::Takagi_mod}), keeping derivatives of $\delta_{dyn}(z,\Delta\theta)$ and approximating $E^{'}_0 (z, \Delta\theta) \simeq 1$ otherwise, we obtain
\begin{subequations}
	\label{dynamical::dynamical_phase}
	\begin{gather}
	\frac{d\delta_{dyn}}{dz} = \left(\frac{k}{2\gamma_0}\right) C \chi_{\overline{h}} e^{i\textbf{hu}(z)}e^{-W(z)}e^{iQz} E^{'}_h (z) \ ,
	\label{dynamical::dynamical_phase_1} \\
	\frac{dE^{'}_{h}}{dz} = \left(\frac{ik}{2\gamma_h} \right) C \chi_{h} e^{-i\textbf{hu}(z)}e^{-W(z)}e^{-iQz}    \ .
	\label{dynamical::dynamical_phase_2}
	\end{gather}
\end{subequations}
The first equation should be complemented with a boundary condition for a dynamical correction $\delta_{dyn}(z=0,\Delta\theta)=0$.

The second equation (\ref{dynamical::dynamical_phase_2}) can be easily calculated leading to well known expression for the diffracted wave in kinematical approximation (compare to results in Ref. \cite{VartKov})
\begin{equation}
E^{'}_h (z, \Delta\theta) = i \left(\frac{kC}{2\gamma_h} \right) \chi_{h} \int_0^z e^{-i\textbf{hu}(z)}e^{-W(z)}e^{-iQz} dz \ .
\label{kin_diffr_wave}
\end{equation}
If the strain field $\textbf{u}(z)$ and the profile of the static Debye-Waller factor $W(z)$ are known then integration can be performed using this equation.
Unfortunately, these parameters commonly are not known and have to be found using other methods.
Especially simple result is obtained in the case of a perfect crystal when $\textbf{u}(z) = W(z) = 0$.
In this case Eq. (\ref{kin_diffr_wave}) reduces to
\begin{equation}
E^{'}_h (z, \Delta\theta) = i \left(\frac{kC}{2\gamma_h} \right) \chi_{h} \int_0^z e^{-iQz} dz \
\label{kin_diffr_wave_1}
\end{equation}
and its integration gives well known expression for the diffracted wave in kinematical approximation for a perfect crystal of thickness $d$
\begin{equation}
E^{'}_h (d, \Delta\theta) = i \left(\frac{Ckd}{2\gamma_h} \right)  \chi_{h} e^{-i\Omega} \left( \frac {\sin\Omega}{\Omega} \right) \ ,
\label{kin_diffr_wave_2}
\end{equation}
where the dimensionless parameter
\begin{equation}
\Omega (\Delta\theta) = Q(\Delta\theta)d/2 = \frac{kd}{4}\frac{\alpha\beta+\chi_0(1-\beta)}{\gamma_0} \approx \frac{kd}{4}\frac{\alpha}{\gamma_h}
\label{omega_def}
\end{equation}
is introduced.

The dynamical contribution to the transmitted wave $\delta_{dyn}(z,\Delta\theta)$ at the exit surface $z=d$ can be obtained by a formal integration of Eq. (\ref{dynamical::dynamical_phase_1})
\begin{equation}
\delta_{dyn}(z,\Delta\theta) = \left(\frac{kC}{2\gamma_0}\right)\chi_{\overline{h}} \int_0^d e^{i\textbf{hu}(z)}e^{-W(z)}e^{iQz} E^{'}_h (z) dz \ ,
\label{nearkin_phase}
\end{equation}
where kinematical solution for the diffracted wave (\ref{kin_diffr_wave}) have to be used.
This is a general expression including also strain fields in the crystal.
It is clear from this expression that the presence of strain can modify the phase of the transmitted beam measured in experiment.
Again, we obtain significant simplification if we consider perfect crystal.
In this case $\textbf{u}(z) = W(z) = 0$, substituting the expression for the diffracted wave $E^{'}_h (z)$ (\ref{kin_diffr_wave_2}) in Eq. \eqref{nearkin_phase} and
performing integration we obtain, finally, for the dynamical correction in the quasi-kinematical approximation
\begin{equation}
\delta_{dyn} (d, \Delta\theta) = - \frac{C^2 \chi_h \chi_{\overline{h}} (kd)^2}{8 \gamma_0 \gamma_h}
                                \frac{1}{\Omega} \left[1 -  e^{i\Omega} \left( \frac{\sin{\Omega }}{\Omega} \right) \right] \ .
\label{nearkin_phase_1}
\end{equation}

Now introducing extinction length $L_{ex}$ \eqref{dynamical::extinction_length} equation \eqref{nearkin_phase_1} can be written in the following form
\begin{equation}
\delta_{dyn} (d, \Delta\theta) =  - \frac{ C_1^2}{2} \left(\frac{d}{L_{ex}}\right)^2
 \frac{1}{\Omega} \left[1 -  e^{i\Omega} \left( \frac{\sin{\Omega }}{\Omega} \right) \right] \ ,
\label{nearkin_phase_2}
\end{equation}
where parameter $C_1$ is introduced after Eq. \eqref{dynamical::scattering_amplitude_definition} in the main text and in the case of a perfect crystal is equal to $C_1 = C(1-ip)$.

Dynamical phase correction in the transmitted beam $\delta\varphi_{dyn} (d, \Delta\theta)$ is given by the real part of this expression

\begin{equation}
\begin{gathered}
\delta\varphi_{dyn} (d, \Delta\theta) = Re\left[  \delta_{dyn} (d, \Delta\theta) \right] =
\\
 = - \frac{ C^2}{2} \left(\frac{d}{L_{ex}}\right)^2
  \frac{1}{\Omega} \left\{ \left( 1-p^2 \right)
 \left[1 -  \cos{\Omega} \left( \frac{\sin{\Omega }}{\Omega} \right) \right]
 - 2p \sin{\Omega} \left( \frac{\sin{\Omega}}{\Omega} \right) \right\} \ .
\label{nearkin_phase_3}
\end{gathered}
\end{equation}

Comparison of expressions \eqref{nearkin_phase_2} and \eqref{kin_diffr_wave_2} shows that the dynamical correction to the transmitted amplitude can be also expressed through the diffracted wave as
\begin{equation}
\delta_{dyn} (d, \Delta\theta) = - \frac{ C_1^2}{2} \left(\frac{d}{L_{ex}}\right)^2
\frac{1}{\Omega} \left[1 + i K e^{2i\Omega}  E^{'}_h (d, \Delta\theta) \right] \ ,
\label{nearkin_phase_4}
\end{equation}
where complex parameter $K=2 \gamma_h / C (kd)\chi_h$ is introduced.

\section{Analytical analysis of the quasi-kinematical approximation}

The derived expression for $\delta_D$ \eqref{approx::phase} is particularly convenient for obtaining analytical results.
First, we should define for which crystal thicknesses the quasi-kinematical approximation is valid.
Since the original assumption of the approximation is $\left| \delta_{dyn}(z ,\Delta\theta) \right| \ll 1$, it is necessary to understand when this condition is satisfied.
Taking a square modulus of the expression \eqref{approx::phase} we obtain

\begin{equation}
	\begin{gathered}
		\label{B2:abs}
		\left|\delta_{dyn} (d, \Delta\theta)\right|^2 =  \frac{C^4 |\chi_h \chi_{\overline{h}}|^2 (kd)^4}{64 \gamma_0^2 \gamma_h^2}
		\frac{1}{\Omega^2} \left|1 -  e^{i\Omega} \left( \frac{\sin{\Omega }}{\Omega} \right) \right|^2 \ =
		\\
		=\ \frac{C^4 |\chi_h \chi_{\overline{h}}|^2 (kd)^4}{64 \gamma_0^2 \gamma_h^2}
		\frac{1}{\Omega^2} \left(1 -  \frac{\sin{2\Omega}}{\Omega} - \frac{\cos{2\Omega}}{2\Omega^2} + \frac{1}{2\Omega^2}\right) \
		.
	\end{gathered}
\end{equation}

Where $\Omega$ is approximated to be real according to expression \eqref{omega_def}.

In the extrema of $|\delta_{dyn}(\Delta\theta)|$, the following condition should be satisfied $d|\delta_D|^2/d\Omega=0$.
This is equivalent to the following equation

\begin{gather}
	\label{B2:globalex}
	\left(\cos{\Omega}-\frac{\sin{\Omega}}{\Omega}\right)^2=0
	.
\end{gather}

The solutions of Eq. \eqref{B2:globalex} are $\Omega_0 = \tan{\Omega_0}$.
Substituting this expression into Eq. \eqref{B2:abs}, we obtain for $\left|\delta_{dyn} (d)\right|^2$ at each extremum

\begin{gather}
\label{B2:globalex_abs}
\left|\delta_{dyn} (d)\right|^2 = \frac{C^4 |\chi_h \chi_{\overline{h}}|^2 (kd)^4}{64 \gamma_0^2 \gamma_h^2} \frac{1}{1+\Omega_0^2}
.
\end{gather}

It follows that the global maximum is present at $\Omega_0 = 0$, and

\begin{gather}
\label{B2:globalex_abs_2}
\left|\delta_{dyn} (d)\right|_{\Omega = 0} = \frac{C^2 |\chi_h \chi_{\overline{h}}| (kd)^2}{8 \gamma_0 \gamma_h}
.
\end{gather}

Substituting expression \eqref{B2:globalex_abs_2} into $\left| \delta_{dyn}(z ,\Delta\theta) \right| \ll 1$, we obtain

\begin{gather}
\label{B2:validity_2}
d \ll \frac{\sqrt{2}}{|C_1|} L_{ex}
.
\end{gather}

Therefore, as one could expect, the near-kinematical approximation is valid for small thicknesses.

Next, we want to find the extrema of the quasi-kinematical expression \eqref{approx_phase_3} for the phase.
Substituting expression \eqref{approx_phase_3} into the condition $d(\delta\varphi_{dyn})/d\Omega=0$, we immediately obtain

\begin{gather}
\left[(1-p^2)\cos{\Omega}+2p\sin{\Omega}\right]\left(\frac{\sin{\Omega}}{\Omega}-\cos{\Omega}\right) = 0.
\label{nearkin_phase_ex}
\end{gather}

The factor on the right gives local extrema for the phase, located at the same points as the extrema of $|\delta_{dyn}(\Delta\theta)|$.
The factor on the left gives

\begin{gather}
\tan{\Omega} = \frac{p^2-1}{2p} = \tan{\left(\frac{1}{\pi} - \varphi_{\chi_h\chi_{\overline{h}}}\right)},
\label{nearkin_phase_ex_omtan}
\end{gather}
where $\varphi_{\chi_h\chi_{\overline{h}}}$ is the complex phase of $\chi_h\chi_{\overline{h}}$.
We are interested in two extrema closest to $\Omega = 0$, which are

\begin{gather}
\Omega_0 = \left(\frac{\pi}{2}-\varphi_{\chi_h\chi_{\overline{h}}} \right)\text{mod}\pi; \left(\frac{\pi}{2}-\varphi_{\chi_h\chi_{\overline{h}}} \right)\text{mod}\pi-\pi,
\label{nearkin_phase_ex_om}
\end{gather}
where $x\ \text{mod}\ y$ represents the nonnegative remainder when dividing $x$ by $y$.

Substituting those $\Omega$ values into Eq. \eqref{approx_phase_3}, we get the extremum value

\begin{gather}
\delta\varphi_{dyn} (d)
= - \frac{ C^2}{2} \left(\frac{d}{L_{ex}}\right)^2 (1-p^2)
\frac{1}{\Omega_0} \ .
\label{nearkin_phase_ex_itself}
\end{gather}

For reflections considered in this work $\varphi_{\chi_h\chi_{\overline{h}}} = 0$, and $\Omega = \pm \pi /2$.
However in crystals without central symmetry $\varphi_{\chi_h\chi_{\overline{h}}} \ne 0$.
The Eq. \eqref{nearkin_phase_ex_itself} shows that the phase $\delta\varphi_{dyn} (d, \Delta\theta)$ modulations indeed grow as a second power of a ratio of a crystal thickness to extinction length $L_{ex}$.

\eject


\begin{table}
	\caption{\label{tab::parameters} Parameters used in simulations presented in Fig. 3 and Fig. 4 for Au and Si crystals. In all simulations $\sigma-$polarization for x-rays was considered}
	\begin{tabular}{l|c|c}
	    Crystal (reflection)                   & Au (111)             & Si (111)           \\
		Energy (keV)                           & 8.5                  & 8.5                \\
        Bragg angle, $\theta_B$ (degrees)      & 18.04                & 13.45              \\
		$\gamma_0$                             & 0                    & 0               \\
		$\gamma_h$                             & 0.808                & 0.892            \\
		$\chi_{0r}$                            & -8.31\ 10$^{-5}$     & -1.35\ 10$^{-5}$  \\
        $\chi_{0i}$                            & 6.87\ 10$^{-6}$      &  2.82\ 10$^{-7}$ \\
        $\chi_{hr}$                            & -6.83\ 10$^{-5}$     & -7.16\ 10$^{-6}$ \\
        $\chi_{hi}$                            & 6.83\ 10$^{-6}$      & 1.97\ 10$^{-7}$ \\
 		$X_r$                                  &  -6.83\ 10$^{-5}$    & -7.16\ 10$^{-6}$          \\
        $X_i$                                  &   6.83\ 10$^{-6}$    & 1.97\ 10$^{-7}$          \\
        $p=-X_i/X_r$                           &   0.10               & 0.027         \\
        Extinction length, $L_{ex}$ (nm)       & 610                  & 6100            \\
        Crystal thickness, $0.2\ L_{ex}$ (nm)  & 120                  & 1200  \\
        Crystal thickness, $0.4\ L_{ex}$ (nm)  &  240                 & 2400 \\
        Crystal thickness, $ 0.6\ L_{ex}$ (nm) & 370                  &  3700 \\
        Crystal thickness, $1.5\ L_{ex}$ (nm)  & 920                  & 9200 \\
	\end{tabular}
\end{table}

\begin{table}
	\caption{\label{tab::parameters_2} Parameters used in simulations presented in Fig. 5 for different crystal reflections}
	\begin{tabular}{l|c|c|c}
		Crystal (reflection)                  & Au (111)              & Au (220)          & Au (222)  \\
		Energy (keV)                           & 8.5                  & 8.5               & 8.5       \\
        Bragg angle, $\theta_B$ (degrees)      & 18.04                & 30.39             & 38.28      \\
		$\gamma_0$                             & 0                    & 0                 & 0           \\
		$\gamma_h$                             & 0.808                & 0.488             & 0.232       \\
		$\chi_{0r}$                            & -8.31\ 10$^{-5}$     & -8.31\ 10$^{-5}$  & -8.31\ 10$^{-5}$      \\
        $\chi_{0i}$                            & 6.87\ 10$^{-6}$      & 6.87\ 10$^{-6}$   & 6.87\ 10$^{-6}$     \\
        $\chi_{hr}$                            & -6.83\ 10$^{-5}$     & -5.63\ 10$^{-5}$  & -5.04\ 10$^{-5}$ \\
        $\chi_{hi}$                            & 6.83\ 10$^{-6}$      & 6.75\ 10$^{-6}$   & 6.69\ 10$^{-6}$ \\
 		$X_r$                                  & -6.83\ 10$^{-5}$     & -5.63\ 10$^{-5}$  & -5.04\ 10$^{-5}$       \\
        $X_i$                                  & 6.83\ 10$^{-6}$      & 6.75\ 10$^{-6}$   & 6.69\ 10$^{-6}$    \\
        $p=-X_i/X_r$                           & 0.10                 & 0.12              & 0.13       \\
        Extinction length, $L_{ex}$ (nm)       & 610                  & 580               & 440      \\
        Crystal thickness, (nm)                & 300                  & 300               & 300  \\		
	\end{tabular}
\end{table}

\begin{table}
	\caption{\label{tab::parameters_3} Parameters used in simulations presented in Fig. 5 for different incoming photon energies}
	\begin{tabular}{l|c|c|c}
		Crystal (reflection)                  & Au (111)              & Au (111)          & Au (111)  \\
		Energy (keV)                          & 5                     & 8.5               & 12        \\
        Bragg angle, $\theta_B$ (degrees)     & 31.77                 & 18.04             & 12.67   \\
		$\gamma_0$                            & 0                     & 0                 & 0         \\
		$\gamma_h$                            & 0.44                  & 0.808             & 0.904         \\
		$\chi_{0r}$                           & -2.40\ 10$^{-4}$      & -8.31\ 10$^{-5}$  & -3.806\ 10$^{-5}$      \\
        $\chi_{0i}$                           &  4.40\ 10$^{-5}$      &  6.87\ 10$^{-6}$  &  5.03\ 10$^{-6}$     \\
        $\chi_{hr}$                           & -1.97\ 10$^{-4}$      & -6.83\ 10$^{-5}$  & -3.07\ 10$^{-5}$  \\
        $\chi_{hi}$                           &  4.37\ 10$^{-5}$      &  6.83\ 10$^{-6}$  &  5.00\ 10$^{-6}$   \\
 		$X_r$                                 & -1.97\ 10$^{-4}$      & -6.83\ 10$^{-5}$  & -3.07\ 10$^{-5}$       \\
        $X_i$                                 &  4.37\ 10$^{-5}$      &  6.83\ 10$^{-6}$  & 5.00\ 10$^{-6}$    \\
        $p=-X_i/X_r$                          &  0.22                 & 0.10              & 0.16       \\
        Extinction length, $L_{ex}$ (nm)      &  270                  & 610               & 1000      \\
        Crystal thickness, (nm)               & 300                   & 300               & 300  \\		
	\end{tabular}
\end{table}

\eject

\eject


\eject

\eject

\eject

\eject

\eject

\eject

\end{document}